\documentclass[journal=jacsat,manuscript=article]{achemso}
\usepackage{stix}
\usepackage[version=3]{mhchem}
\usepackage{soul}
\usepackage{siunitx}
\usepackage{physics}
\usepackage{multirow}
\usepackage{xcolor}
\usepackage{amsmath}
\usepackage{bm}
\usepackage{xr}

\usepackage{hyperref}
\usepackage{setspace}
\usepackage{tcolorbox}
\usepackage{soul}

\newcommand{\eql}[1]{\label{eq:#1}}

\newcommand{\bse}{\begin{subequations}}
\newcommand{\ese}{\end{subequations}}

\usepackage{etoolbox}

\makeatletter
\renewcommand*{\acs@author@fnsymbol@symbol}[1]{
    \ifcase #1 *\or
    1\or
    2\or
    3\or
    4\or
    5\or
    6\or
    7\or
    8\or
    9\or
    10
    \fi
}
        
\renewcommand*\acs@contact@details{
    {\sffamily *\,E-mail: \acs@email@list }%
    \acs@number@list
}           

\patchcmd{\acs@address@list@auxii}
{\acs@author@fnsymbol{\acs@affil@marker@cnt}}
{\textsuperscript{\acs@author@fnsymbol{\acs@affil@marker@cnt}}}
{}{}

\patchcmd{\acs@address@list@auxii}
{{\acs@author@fnsymbol{\acs@affil@marker@cnt}\@nameuse{@altaffil@\@roman\@tempcnta}\par}}
{{\textsuperscript{\acs@author@fnsymbol{\acs@affil@marker@cnt}}\@nameuse{@altaffil@\@roman\@tempcnta}\par}}
{}{}        

\makeatother

\author{Lujo Matasovic}
\affiliation{Cavendish Laboratory, University of Cambridge, Cambridge, UK}
\author{Petri Murto}
\affiliation{Cavendish Laboratory, University of Cambridge, Cambridge, UK}
\alsoaffiliation{Yusuf Hamied Department of Chemistry, University of Cambridge, Cambridge, UK}
\alsoaffiliation{Department of Chemistry and Materials Science, Aalto University, Espoo, Finland}
\author{Shilong Yu}
\affiliation{State Key Laboratory of Supramolecular Structure and Materials, Jilin University, Changchun, P. R. China}
\author{Wenzhao Wang}
\affiliation{State Key Laboratory of Supramolecular Structure and Materials, Jilin University, Changchun, P. R. China}
\author{James D. Green}
\affiliation{Department of Chemistry, University College London, London, UK}
\author{Giacomo Londi}
\affiliation{Department of Chemistry and Industrial Chemistry, University of Pisa, Pisa, Italy}
\author{Weixuan Zeng}
\affiliation{Yusuf Hamied Department of Chemistry, University of Cambridge, Cambridge, UK}
\author{Laura Brown}
\affiliation{Yusuf Hamied Department of Chemistry, University of Cambridge,  Cambridge, UK}
\author{William K. Myers}
\affiliation{Centre for Advanced ESR, Department of Chemistry, University of Oxford, Oxford, UK}
\author{David Beljonne}
\affiliation{Laboratory for Chemistry of Novel Materials, University of Mons, Mons, Belgium}
\author{Yoann Olivier}
\affiliation{Laboratory for Computational Modelling of Functional Materials, University of Namur, Namur, Belgium}
\author{Feng Li}
\email{lifeng01@jlu.edu.cn}
\affiliation{State Key Laboratory of Supramolecular Structure and Materials, Jilin University, Changchun, P. R. China}
\author{Hugo Bronstein}
\email{hab60@cam.ac.uk}
\affiliation{Yusuf Hamied Department of Chemistry, University of Cambridge, Cambridge, UK}
\author{Timothy J. H. Hele}
\email{t.hele@ucl.ac.uk}
\affiliation{Department of Chemistry, University College London, London, UK}
\author{Richard H. Friend}
\email{rhf10@cam.ac.uk}
\affiliation{Cavendish Laboratory, University of Cambridge,  Cambridge, UK}
\author{Sebastian Gorgon}
\email{sg911@cam.ac.uk}
\affiliation{Cavendish Laboratory, University of Cambridge, Cambridge, UK}
\alsoaffiliation{Centre for Advanced ESR, Department of Chemistry, University of Oxford, Oxford, UK}

\title[]
   {Coulombic Control of Charge Transfer in Luminescent Radicals with Long-Lived Quartet States}
   
\begin{document}
\newpage
\begin{abstract}
\noindent
\textbf{Excitons in organic materials are emerging as an attractive platform for tunable quantum technologies. Structures with near-degenerate doublet and triplet excitations in linked trityl radical, acene and carbazole units can host quartet states. These high spin states can be coherently manipulated, and later decay radiatively via the radical doublet transition. However, this requires controlling the deexcitation pathways of all metastable states. Here we establish design rules for efficient quartet generation in luminescent radicals, using different connection arrangements of the molecular units. We discover that electronic coupling strength between these units dictates luminescence and quartet formation yields, particularly through the energetics of an acene-radical charge transfer state, which we tune Coulombically. This state acts as a source of non-radiative decay when acene-radical separation is small, but facilitates doublet-quartet spin interconversion when acene-radical separation is large. Using these rules we report a radical-carbazole-acene material with 55\% luminescence yield, where 94\% of emitting excitons originate from the quartet at microsecond times. This reveals the central role of molecular topology in luminescent quantum materials.}
\end{abstract}

\section{Introduction}

Robust spin-optical interfaces are crucial to harnessing the quantum resources of materials, as light provides a facile way to interact with a system.\cite{Awschalom2018} Most interfaces to date utilize defects in inorganic materials, such as diamond.\cite{Doherty2013} However, the photoluminescence yield of defects is typically low, and the control of interactions between spin centers is challenging. An emerging alternative involves using molecules, offering vast tunability via synthetic modulation.\cite{Atzori2019, Wasielewski2020} Organic materials in particular support multi-spin interactions due to electron delocalisation via $\pi$-pathways.\cite{Teki1996} Incorporating stable radicals is attractive since these can contribute to states active in Electron Spin Resonance (ESR), both in the ground (doublet, $S=1/2$) and excited (quartet, $S=3/2$) state.\cite{Quintes2023} Quartets rely on engineering exchange coupling between a pair of triplet spins and the third spin on the pendant radical.\cite{Kollmar1982} Critically, such compounds have thus far been non-luminescent,\cite{Teki2006, Giacobbe2009} limiting their potential. We recently utilized a derivative of the tris(2,4,6-trichlorophenyl)- methyl (TTM) radical to engineer a luminescent pathway following quartet generation.\cite{Gorgon2023} This was enabled by the doublet emission channel in substituted TTM radicals,\cite{Matsuoka2022,Mizuno2024} which have previously been utilized in organic light emitting diodes with record efficiencies.\cite{Ai2018,Gu2024,Cho2024} 
As only a singular design of luminescent TTM radical-chromophore dyads is currently known, the structure-property relationship remains unclear.

In this work we establish general strategies for controlling quartet state generation in luminescent radicals. We synthesized new compounds, with three distinct molecular building blocks available for tuning: a TTM core (T), a carbazole, and an acene unit with a triplet energetically close to the emissive state (Fig. \ref{fgr:systems}a). An-T-1Cz and An-T-3PCz show strong radical-acene electronic coupling, as the anthracene is connected directly to the radical. The carbazole bridges the units in T-3Cz-An and T-3Cz-Acr, and these compounds show weak radical-acene electronic coupling. Acridine was chosen as it has a higher triplet energy (1.96 eV) compared to anthracene (1.82 eV).\cite{Ford1991} We identify four excited states required for their operation (Fig. \ref{fgr:systems}d): A bright carbazole-to-TTM charge-transfer state ($^{2}$CT$_{\text{Cz}}$); two local excitations on the triplet-bearing unit ($^{2}$LE$_{\text{Ac}}$ and $^{4}$LE$_{\text{Ac}}$), and an acene-to-TTM charge-transfer state ($^{2}$CT$_{\text{Ac}}$). The latter is crucially important for the performance of these materials as we discover below.

\section{Results}
\subsection{Optical Properties}

\begin{figure}[t!]
    \centering
    \includegraphics[width=1\textwidth]{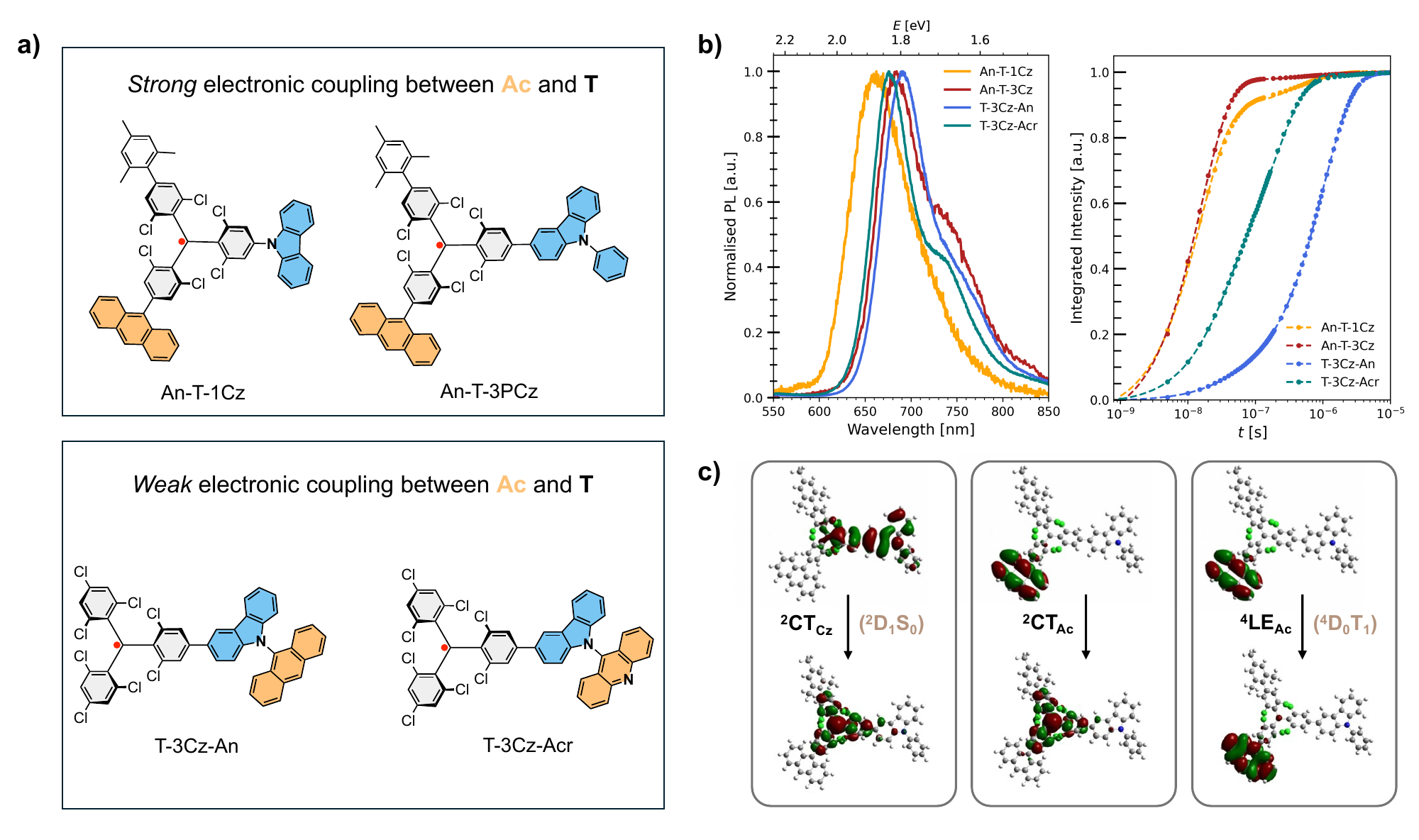}
  \caption{\textbf{Effects of topology in radical-acene dyads.} a) Chemical structures. In An-T-1Cz and An-T-3PCz, the anthracene and carbazole are both coupled directly to the radical, providing strong radical-acene electronic coupling. The radical is decorated with mesityl groups providing stabilization via extra steric protection without affecting the photophysics.\cite{Murto2023} In T-3Cz-An and T-3Cz-Acr, the acene is linked to the radical via carbazole, providing weak radical-acene electronic coupling. b) Normalized steady-state emission spectra. c) Normalized integrated transient PL curves. Experimental data shown in dots, biexponential fits shown in dashed lines. PL experiments performed in 0.1 mM toluene solutions at room temperature under band gap excitation. T-An is non emissive. d) Hole-particle natural transition orbitals (NTOs) of An-T-3PCz, corresponding to optical transitions involved in quartet generation and optical readout: $^{2}$CT$_{\text{Cz}}$ is the luminescent doublet CT state, $^{2}$CT$_{\text{Ac}}$ the dark CT state involving a transition between the TTM and acene and $^{2,4}$LE$_{\text{Ac}}$ doublet/quartet locally excited state on the acene unit.}
  \label{fgr:systems}
\end{figure}

Optical properties were measured at room temperature in 100 µM toluene solution, unless stated otherwise. The four reported compounds are luminescent (Fig. \ref{fgr:systems}b), as summarized in Table \ref{tab:photoPar}. Materials with strong radical-acene electronic coupling have modest photoluminescence quantum yields (PLQY), with 3\% in An-T-3PCz and 8\% in An-T-1Cz. In contrast, materials with weak radical-acene electronic coupling reach emission efficiencies of 55\% in T-3Cz-An and 50\% in T-3Cz-Acr. This stark difference cannot be explained by the existing understanding of luminescent high-spin radicals,\cite{Gorgon2023} as both groups fulfill the energetic resonance condition. Additionally, we study the carbazole-free T-An as a control. This compound shows absorption similar to M$_{3}$TTM (Fig. S10),\cite{Murto2023} but it is not emissive, as expected due to the 0.3 eV energy gap between the doublet emission and the triplet of anthracene.\cite{Bredas2025-triplets}

\begin{table}[bt]
    \centering
    \begin{tabular}{lrrrS[table-format=2.1]rS[table-format=4.2]r}
     Compound & $\lambda_{PL}$ [nm] & $\phi_{PL}$ [\%] & {$\tau_1$ [ns]} & {$\phi_{1}$ [\%]} & {$\tau_2$ [ns]} & {$\phi_{2}$ [\%]} & {$\Delta E$[meV]} \\ 
    \hline\hline
     An-T-1Cz & 662 & 8 & 15.6 & 7.1 & 437 & 0.9 & $23 \pm 4$ \\
     An-T-3PCz & 684 & 3 & 15.8 & 2.9 & 729 & 0.06 & $21 \pm 6$ \\ 
    \hline
     T-3Cz-An & 690 & 55 & 31.6 & 3.3 & 1030 & 51.7 & $15 \pm 4$ \\
     T-3Cz-Acr & 676 & 50 & 31.8 & 18.0 & 216 & 32.0 & $11 \pm 4$ \\ 
                                 
    \hline\hline
\end{tabular}
    \caption{\textbf{Photophysical characterization.} Photoluminescence peak wavelength ($\lambda_{PL}$), quantum efficiency ($\phi_{PL}$), prompt ($\tau_1$) and delayed ($\tau_2$) emission lifetimes with their contributions to the quantum yield ($\phi_{1}+\phi_{2}=\phi_{PL}$) measured in 0.1 mM toluene solutions at room temperature. Effective activation energies ($\Delta E$) were determined from temperature-dependent transient PL measurements in 5 wt\% PMMA films.}
    \label{tab:photoPar}
\end{table}

The photoluminescence (PL) provides a powerful probe of excited-state dynamics. All emissive compounds exhibit bi-exponential decay kinetics in solution (Fig. \ref{fgr:systems}c and S11-12), but the delayed emission rates vary significantly. We calculated quantum yields of the two emission components, $\phi_{1}$ and $\phi_{2}$, taken as the product of PLQY, $\phi_{PL}$, and the relevant fitted weight. Delayed emission dominates only in compounds with weak radical-acene electronic coupling. As this follows quartet-doublet reverse intersystem crossing (RISC), the dominance of this channel is desired to allow time for quartet addressability. The quartet is depleted 5 times more rapidly in T-3Cz-Acr than T-3Cz-An, making the latter the champion compound, with up to threefold improvement on the previously reported $\phi_{2} =$ 18\%.\cite{Gorgon2023}

As CT states are sensitive to environment polarity, we study T-3Cz-An in more polar tetrahydrofuran (Fig. S13). Emission is weaker and faster compared to toluene, with prompt emission becoming dominant (55\%), with lifetimes of $\tau_{1}$ = 16.7 ns and $\tau_{2}$ = 136 ns. This indicates that the excited state energy reordering in a more polar solvent predominantly hinders the delayed emission. We turn to temperature-dependent time-resolved PL measurements on films at 5\% radical loading into poly(methyl methacrylate) (PMMA) to investigate the energetic landscape around the emissive state. As the molecules are conformationally locked in a rigid medium, we observe a small gradual hypsochromic shift in all molecules (Fig. S14-S15). The effective lifetime extends to around 2.3 µs for films of T-3Cz-An. In contrast, it shortens in films of An-T-3PCz, indicating the promotion of deactivation channels in a more rigid environment.\cite{Yamaji2021} Using an Arrhenius-type model, we calculate the activation energies of delayed emission, which are below $k_{\text{B}}T$ at room temperature (Fig. S16-17) for all materials. 

\begin{figure}[b!]
    \centering
    \includegraphics[width=1\textwidth]{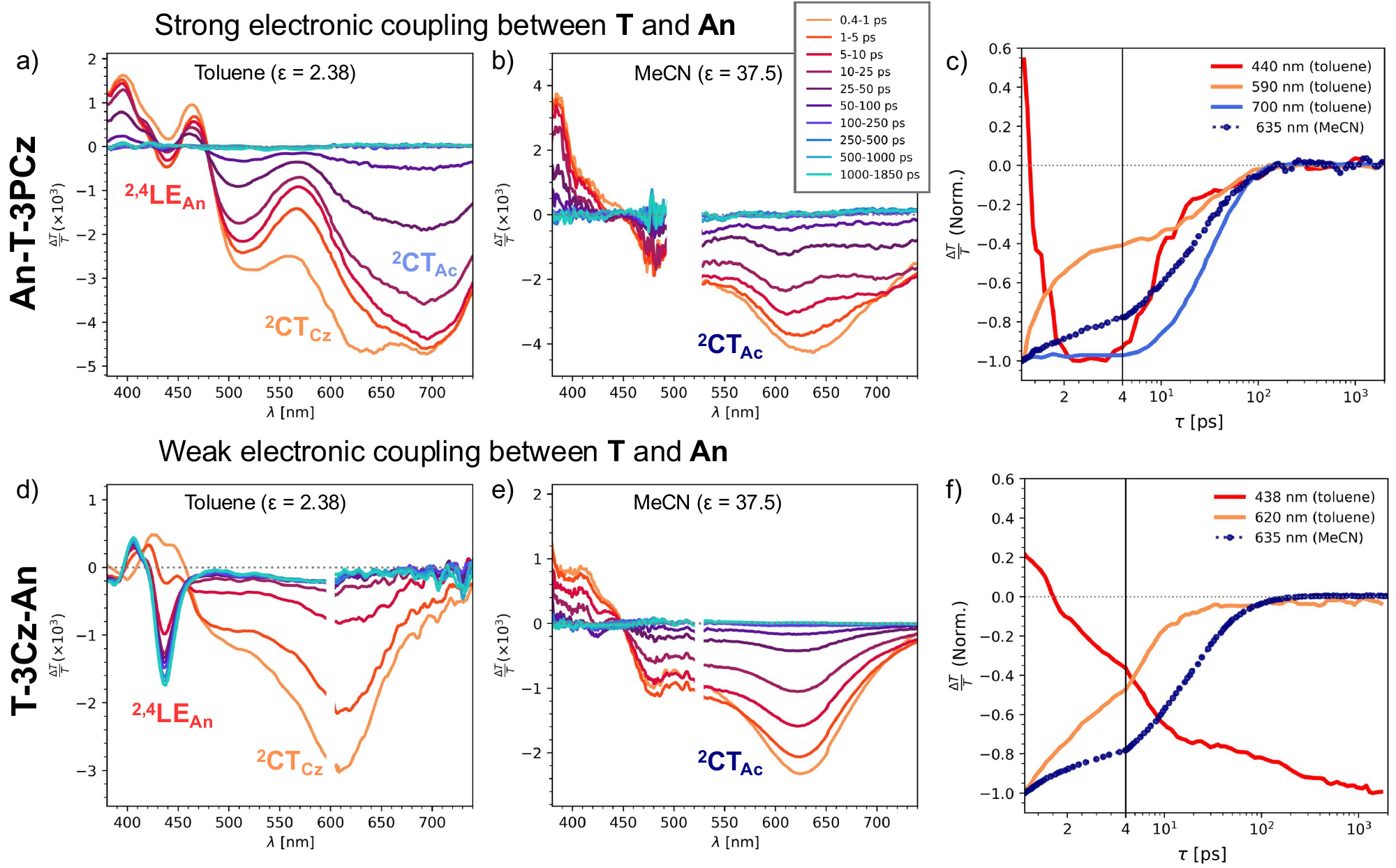}
  \caption{\textbf{Dielectric control of non-radiative decay.} Transient absorption spectral slices and corresponding kinetic profiles in toluene and acetonitrile (MeCN) for An-T-3PCz (a-c) T-3Cz-An (d-f). Color-coding of time slices is consistent across the panels. Extracted kinetics in toluene reveal sub-10 ps wavefunction localization onto the anthracene unit in both compounds.}
  
  \label{fgr:ta_trpl}
\end{figure}

We use Transient Absorption (TA) spectroscopy to track the complete excited state dynamics of our materials across ps-µs times (Fig. 2 and S18). In dilute toluene solutions, the prompt spectrum includes a ground-state bleach (GSB) in the UV region, a D$_1$ photoinduced absorption (PIA) around 600 nm, and a broad IR PIA due to CT states. The carbazole-containing molecules have strong IR bands above 1600 nm, characteristic of $^{2}$CT$_{\text{Cz}}$ state.\cite{Ai2018} In An-T-3PCz we also resolve a band at 700 nm, matching the anthracene cation,\cite{Khan1992} which we assign to $^{2}$CT$_{\text{Ac}}$. The growing PIA at 430 nm is due to the anthracene triplet, arising from both $^{2}$LE$_{\text{Ac}}$ and $^{4}$LE$_{\text{Ac}}$ populations. This band is convolved with the GSB to varying extent. 

Strong radical-acene electronic coupling allows rapid interconversion between $^{2}$CT$_{\text{Ac}}$ and $^{2}$LE$_{\text{Ac}}$ states. In An-T-1Cz, An-T-3PCz and T-An, all transient features decay within 100 ps (Fig. S18), indicating most of the decay occurs from a state of similar character. This indicates strong non-radiative coupling of $^{2}$CT$_{\text{Ac}}$ and $^{2}$LE$_{\text{Ac}}$ to the ground state.  
However, for a 5\% An-T-3PCz in PMMA film the spectra are preserved and dynamics are slower compared to measurements in solution (Fig. S20). The signals persist beyond 8 ns, indicating non-radiative decay can be controlled conformationally. For compunds with weak radical-acene electronic coupling, the decay of the initial $^{2}$CT$_{\text{Cz}}$ PIA matches the rise of the $^{2}$LE$_{\text{Ac}}$, with a lifetime of $\tau_{DT}=3.8\pm0.1$ ps for T-3Cz-An and $\tau_{DT}=4.6\pm0.1$ ps for T-3Cz-Acr. We do not observe an accumulation of population in the $^{2}$CT$_{\text{Ac}}$ band. In T-3Cz-An, the 430 nm PIA then decays with a 1.55 µs lifetime in toluene (Fig. S21-22). This matches the delayed emission lifetime, indicating small non-radiative losses at late times. 

Finally, we turn to solvatochromic TA studies (Fig. S19). We observe no significant differences in the $^{2,4}$LE$_{\text{Ac}}$ feature in less polar cyclohexane. In dichloromethane, we still observe this PIA, but its lifetime is quenched (Fig. S22). In acetonitrile (MeCN), T-3Cz-An spectra resemble those of An-T-3PCz, and the dominant $^{2}$CT$_{\text{Ac}}$ band follows the same kinetics in both. The $^{2,4}$LE$_{\text{Ac}}$ PIA is also suppressed, showing that the energetic ordering of states with CT character changed at high polarity. $^{2}$CT$_{\text{Ac}}$ becomes the lowest excited state upon stabilization in all compounds. This stabilization is the largest for compounds with weak radical-acene electronic coupling, due to a large electron-hole separation in $^{2}$CT$_{\text{Ac}}$.

\subsection{Magnetic Properties} 

\begin{figure}[t!]
    \centering
    \includegraphics[width= \textwidth]{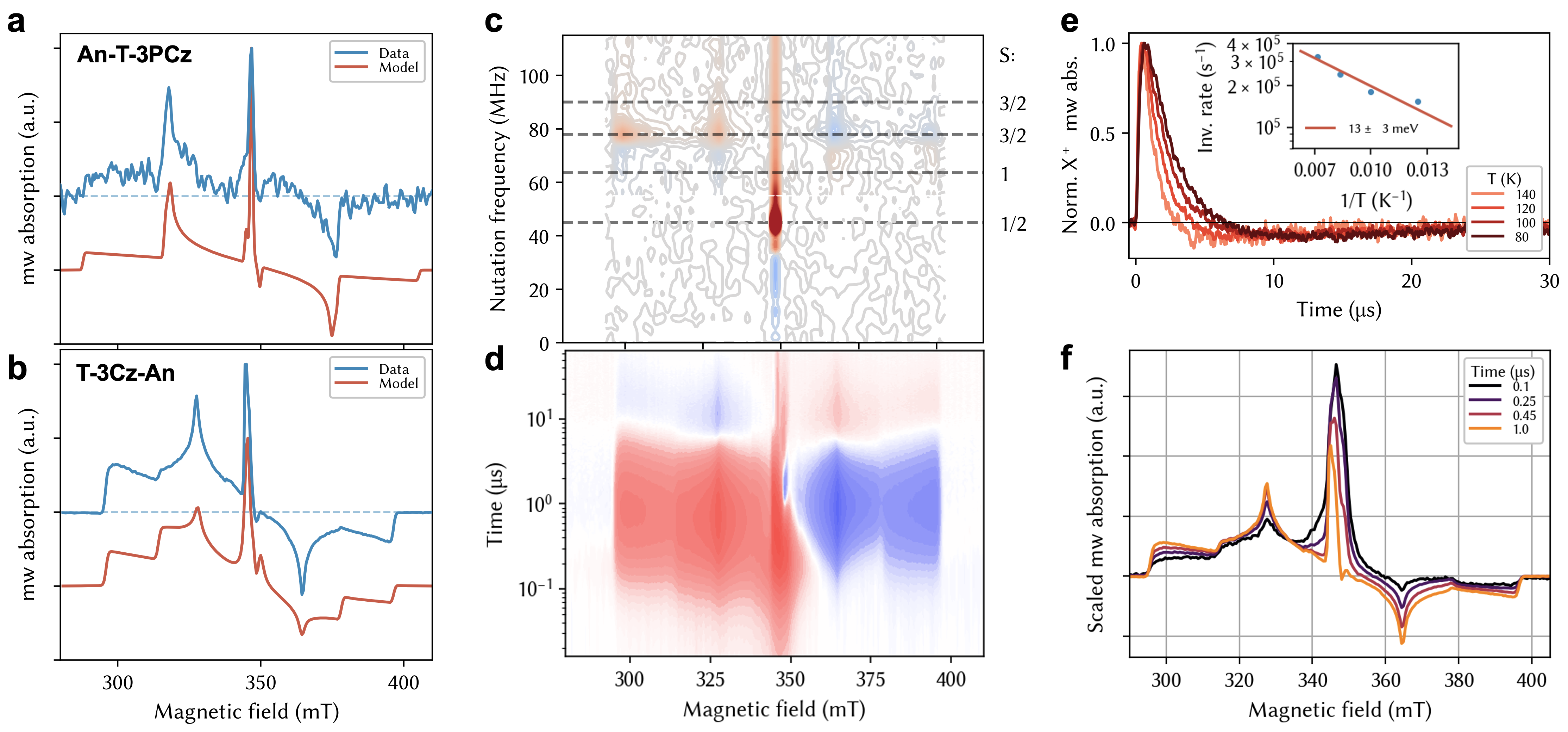}
  \caption{\textbf{Quartet excited state ESR.} Performed on 100 µM toluene frozen solutions at 80 K at X-band. Prompt trESR on  a) An-T-3PCz, and b) T-3Cz-An. c) Transient nutation map confirming spin multiplicity of recorded transitions in T-3Cz-An, indicating the positions calculated based on the dark nutation signal and the state spin quantum number \textit{S}.  d) 2D trESR map in T-3Cz-An. e) Temperature dependence of inversion rate of T-3Cz-An polarization monitored at 328 mT. Inset shows Arrhenius analysis. f) Rise time trESR spectra in T-3Cz-An. A broad absorptive feature is visible at 100 ns, which is buried below the rising quartet polarization. }
  \label{fgr:epr}
\end{figure}

Transient ESR (trESR) performed at X-band reveals the formation of strongly-exchange coupled $^4$LE$_{\text{Ac}}$ states in all four reported compounds (Fig. 3 and S23-24). The sublevel population patterns extracted from the simulations of the prompt spectra show an overpopulation of the inner quartet sublevels,\cite{Tait2023} suggesting a similar intersystem crossing (ISC) mechanism is active regardless of strength of radical-acene electronic coupling (Table 2). We do not detect excited-state signals in the trESR of T-An, indicating that non-radiative relaxation outcompetes ISC. 

The width of the full field spectra is controlled by $D_Q = \frac{1}{3}(D_T+D_{TR})$, where $D_T = 2145$ MHz for anthracene and $D_T = 2208$ MHz for acridine.\cite{Grivet1971} The distance $r_{TR}$ between the triplet, and the third radical spin which makes up the quartet exciton can be estimated from the additional radical-triplet dipolar coupling $D_{TR}$.\cite{Colvin2010} The $r_{TR}$ values correspond well to the separation between the center of anthracene and the central radical carbon, confirming that the $^2$CT$_{\text{Cz}}$ wavefunction does not contribute to the observed quartet states. The dipolar constants ($D_Q$, $E_Q$) reveal the shape of the quartet spin density, with more axial symmetry seen in compounds with direct radical-acene coupling.\cite{Hintze2017} Population inversion occurs with a barrier of $\Delta$E$_{A} = 13\pm3$ meV for T-3Cz-An (Fig. 3e), matching the activation energy for delayed emission.

\begin{table}[b!]
\begin{tabular}{llrrc}

 {Compound} & \multicolumn{1}{c}{{\begin{tabular}[c]{@{}c@{}}$|D_Q|, |E_Q|$ \\ (MHz)\end{tabular}}} & \multicolumn{1}{c}{{\begin{tabular}[c]{@{}c@{}}$|D_{TR}|$\\ (MHz)\end{tabular}}} & \multicolumn{1}{c}{{\begin{tabular}[c]{@{}c@{}}$r_{TR}$\\ (nm)\end{tabular}}} & \begin{tabular}[c]{@{}l@{}}Populations at 1 µs\\ $Q_{+3/2}$: $Q_{+1/2}$:$Q_{-1/2}$:$Q_{-3/2}$:$D_{+1/2}$:$D_{-1/2}$\end{tabular} \\
\hline\hline
 An-T-1Cz  & 820, -10 & 315 & 0.63 & 0.00 : 0.41 : 0.47 : 0.12 : 0.00 : 0.23\\
 An-T-3PCz & 835, -11 & 360 & 0.60 & 0.18 : 0.29 : 0.35 : 0.18 : 0.00 : 0.10\\\hline
 T-3Cz-An  & 690, -90 & 75  & 1.01 & 0.13 : 0.34 : 0.38 : 0.15 : 0.00 : 0.11 \\
 T-3Cz-Acr & 710, -70 & 68  & 1.05 & 0.04 : 0.49 : 0.43 : 0.16 : 0.00 : 0.02 \\\hline
\hline
\end{tabular}
\caption{\textbf{Spin Hamiltonian parameters extracted from trESR spectra.}}
\label{tab:trEPR}
\end{table}

Additionally, in T-3Cz-An we observe a homogeneously broadened absorptive feature centered at $g=2.004$ with width $\Delta\omega\approx5.5$ mT (Fig. 3d). This feature narrows during the rise of the quartet signal (Fig. 3f). Due to similarities with timescales extracted from optical methods we propose it is associated with the overpopulation of the $m_s=-1/2$ sublevel of the $^2$CT$_{\text{Ac}}$ state during ISC.\cite{Fujisawa2001}

The high quartet yield (Fig. S25) in T-3Cz-An allows us to perform pulsed ESR spectroscopy. The phase memory time of the T-3Cz-An $^4$LE$_{\text{Ac}}$ state at 80 K in frozen dilute protonated toluene is $T_{m} = 0.90 $ µs (Fig. S26). To confirm the spin multiplicity across the whole spectrum, we perform a light-induced nutation pulse sequence as a function of field.\cite{Stoll1998} The resulting map (Fig. 3c) confirms that there are no uncoupled triplet excitons in the molecular ensemble.

\subsection{Quantum-Chemical Calculations}

\begin{table}[b!]
\begin{tabular}{llllllllll}
\multicolumn{5}{c|}{\textbf{Toluene}} & \multicolumn{5}{c}{\textbf{MeTHF}} \\ \hline \hline
\multicolumn{1}{r|}{\textbf{\textit{E} [eV]}} & \multicolumn{1}{r|}{\textbf{f}} & \multicolumn{1}{r|}{$\mathbf{R_{\text{he}}}$} & \multicolumn{1}{r|}{$\phi_s$} & \multicolumn{1}{l|}{\textbf{State}}  & \multicolumn{1}{r|}{\textbf{\textit{E} [eV]}} & \multicolumn{1}{r|}{\textbf{f}}  & \multicolumn{1}{r|}{$\mathbf{R_{\text{he}}}$} & \multicolumn{1}{r|}{$\phi_s$} & \textbf{State}  \\ \hline \hline

\multicolumn{10}{c}{\textbf{An-T-3PCz}} \\ \hline
\multicolumn{1}{r|}{2.06}   & \multicolumn{1}{r|}{0.000} & \multicolumn{1}{r|}{0.05} & \multicolumn{1}{r|}{0.89}  & \multicolumn{1}{l|}{$^{2,4}$LE$_{\text{Ac}}$ } & \multicolumn{1}{r|}{2.05}   & \multicolumn{1}{r|}{0.001} & \multicolumn{1}{r|}{1.05} & \multicolumn{1}{r|}{0.83}  & $^{2,4}$LE$_{\text{Ac}}$  \\ \hline
\multicolumn{1}{r|}{2.28}   & \multicolumn{1}{r|}{0.007} & \multicolumn{1}{r|}{6.74} & \multicolumn{1}{r|}{0.21}  & \multicolumn{1}{l|}{$^{2}$CT$_{\text{Ac}}$}  & \multicolumn{1}{r|}{2.11}   & \multicolumn{1}{r|}{0.004} & \multicolumn{1}{r|}{5.90} & \multicolumn{1}{r|}{0.13}   & $^{2}$CT$_{\text{Ac}}$\\ \hline
\multicolumn{1}{r|}{2.41}   & \multicolumn{1}{r|}{0.121} & \multicolumn{1}{r|}{5.58} & \multicolumn{1}{r|}{0.50}  & \multicolumn{1}{l|}{$^{2}$CT$_{\text{Cz}}$} & \multicolumn{1}{r|}{2.26}  & \multicolumn{1}{r|}{0.115} & \multicolumn{1}{r|}{6.78} & \multicolumn{1}{r|}{0.42}  & $^{2}$CT$_{\text{Cz}}$ \\ \hline
\multicolumn{10}{c}{\textbf{T-3Cz-An}} \\ \hline \hline
\multicolumn{1}{r|}{2.06}   & \multicolumn{1}{r|}{0.000} & \multicolumn{1}{r|}{0.06} & \multicolumn{1}{r|}{0.89}  & \multicolumn{1}{l|}{$^{2,4}$LE$_{\text{Ac}}$ } & \multicolumn{1}{r|}{2.05}   & \multicolumn{1}{r|}{0.000} & \multicolumn{1}{r|}{0.06} & \multicolumn{1}{r|}{0.89}  & $^{2,4}$LE$_{\text{Ac}}$  \\ \hline
\multicolumn{1}{r|}{2.34}   & \multicolumn{1}{r|}{0.130} & \multicolumn{1}{r|}{5.78} & \multicolumn{1}{r|}{0.48}  & \multicolumn{1}{l|}{$^{2}$CT$_{\text{Cz}}$} & \multicolumn{1}{r|}{2.14} & \multicolumn{1}{r|}{0.113} & \multicolumn{1}{r|}{7.18} & \multicolumn{1}{r|}{0.39}  & $^{2}$CT$_{\text{Cz}}$ \\ \hline
\multicolumn{1}{r|}{2.76}   & \multicolumn{1}{r|}{0.001} & \multicolumn{1}{r|}{10.54} & \multicolumn{1}{r|}{0.22}   & \multicolumn{1}{l|}{$^{2}$CT$_{\text{Ac}}$}  & \multicolumn{1}{r|}{2.38}  & \multicolumn{1}{r|}{0.007} & \multicolumn{1}{r|}{11.38} & \multicolumn{1}{r|}{0.14}   & $^{2}$CT$_{\text{Ac}}$\\ \hline
\end{tabular}
\caption{\textbf{Excited-state calculations.} Vertical excitation energies (\textit{E}), oscillator strengths (f), electron-hole separation distance ($R_{\text{he}}$) and hole-electron density overlap ($\phi_{s}$) of low-lying transitions, calculated in toluene and MeTHF for representative molecules with strong (An-T-3PCz) and weak (T-3Cz-An) radical-acene electronic coupling.}
\label{tab:dftCalc}
\end{table}

We carried out excited-state calculations both in toluene and 2-methyltetrahydrofuran (MeTHF) by means of time-dependent density functional theory (TDDFT) calculations, as summarized for T-An-3Cz and An-T-3PCz in Table \ref{tab:dftCalc} and Figure \ref{fgr:systems}c. For each system, we identify the key excited states using natural transition orbitals (NTOs, Fig. \ref{fgr:systems}d). In all molecules, the $^{2,4}$LE$_{\text{Ac}}$ and  $^{2}$CT$_{\text{Ac}}$ states show a vanishing oscillator strength, in contrast to the brighter  $^{2}$CT$_{\text{Cz}}$, in agreement with experiments (Tables S4-8). Regardless of the solvent, $^{2,4}$LE$_{\text{Ac}}$ states are below the two CT states for all calculated systems. The lowest-lying CT state in compounds with weak radical-acene electronic coupling is the bright $^{2}$CT$_{\text{Cz}}$. The trend is reversed in molecules with strong radical-acene electronic coupling, where the bright $^{2}$CT$_{\text{Cz}}$ lies above the dark $^{2}$CT$_{\text{Ac}}$. The low-lying $^{2}$CT$_{\text{Ac}}$ state acts as a dark trap, competing both with the generation of quartet states and delayed luminescence.

The state ordering remains the same even in the solvent of higher polarity, but the energy difference between $^{2}$CT$_{\text{Ac}}$ and $^{2}$CT$_{\text{Cz}}$ states of T-3Cz-An is reduced. The stabilization of $^{2}$CT$_{\text{Ac}}$ is more significant than for the $^{2}$CT$_{\text{Cz}}$ state, due to a larger electron-hole separation of $^{2}$CT$_{\text{Ac}}$, as suggested by the calculated hole-electron separation distance $R_{he}$ and by the hole-electron density overlap metric $\phi_{s}$. In contrast, the CT state energy difference in An-T-3PCz is not affected by solvent polarity, as their molecular topology dictates a comparable electron-hole separation in both CT states. We note that, in the case of T-3Cz-Acr, quantum chemical calculations show that the $^{2}$CT$_{\text{Ac}}$ state lies around 0.7 eV above the $^{2}$CT$_{\text{Cz}}$ in toluene. A smaller number of conformers that can populate the $^{2}$CT$_{\text{Ac}}$ at room temperature might explain the larger contribution of prompt emission of T-3Cz-Acr compared to T-3Cz-An.

\subsection{Theoretical Insights}

In this section, we develop the energetic criteria for the radical-acene dyads with a luminescent channel from the quartet state by combining electronic structure algebra for the excited states of radicals with perturbation theory.\cite{Hele2021,Hele2019,Green2022,Ai2018,Abdurahman2020,Li2022} The core results are presented here, and their full derivation can be found in SI Section 5.

For both ISC and RISC to occur efficiently, the quartet state must lie below the emissive radical state by roughly the thermal energy. At zeroth order, if the $D_1$ energy of the radical is similar to the $T_1$ energy of the acene, the bright $^2$D$_1$S$_0$ state will be of similar energy to the quartet state $^4$D$_0$T$_1$ in the dimer. However, at higher levels of theory, the quartet energy of the dimer will drop below the triplet energy of the isolated acene by approximately $\frac{1}{2}(K_{A1,R0}+K_{R0,A1^{\prime}})$, which is half the exchange energy of an electron in the radical SOMO with one in the acene HOMO, plus half the exchange energy of an electron in the radical SOMO with one in the acene LUMO.\cite{Green2024} The RISC is therefore efficient when the sum of this exchange stabilisation and the $D_1$-$T_1$ energy gap is comparable to $kT$. The quartet state $^4$D$_0$T$_1$ should also be energetically stabilised relative to the equivalent doublet state $^2$D$_0$T$_1$. This can be achieved if the magnitude of the exchange stabilisation is large relative to that of the magnetic dipolar coupling. This exchange stabilisation is $ E(^2$D$_0$T$_1)-E(^4$D$_0$T$_1)=\frac{3}{2}(K_{A1,R0}+K_{R0,A1^{\prime}}),\eql{xc}$
and is shown to increase with planarity between the radical and acene.\cite{Green2024} 

The acene-to-radical CT state must be sufficiently above the quartet state in energy, as otherwise it acts as a trap state. This $^2$CT$_{\text{Ac}}$ state is stabilised by the Coulomb attraction $J_{A1,R0}$ between an electron in the radical SOMO and the hole in the acene HOMO, whereas the quartet is stabilised by the Coulombic attraction $J_{A1,A1^\prime}$ of an electron in the acene LUMO and a hole in the acene HOMO. To avoid non-radiative decay via this $^2$CT$_{\text{Ac}}$ state, we wish to minimise $J_{A1,R0}$ by ensuring that the radical SOMO and acene HOMO are spatially separated and maximise $J_{A1,A1^\prime}$ by ensuring that the acene HOMO and LUMO are in the same part of space. Additionally, there should also be a significant coupling between $^2$D$_1$S$_0$ and $^2$D$_0$T$_1$ to facilitate reversible internal conversion. This can be indirectly mediated via a CT state and the magnitude of this interaction is again maximised by planarity between the radical and acene, as evident from equation S30.

\section{Discussion}

\begin{figure}[b!]
    \centering
    \includegraphics[width= \textwidth]{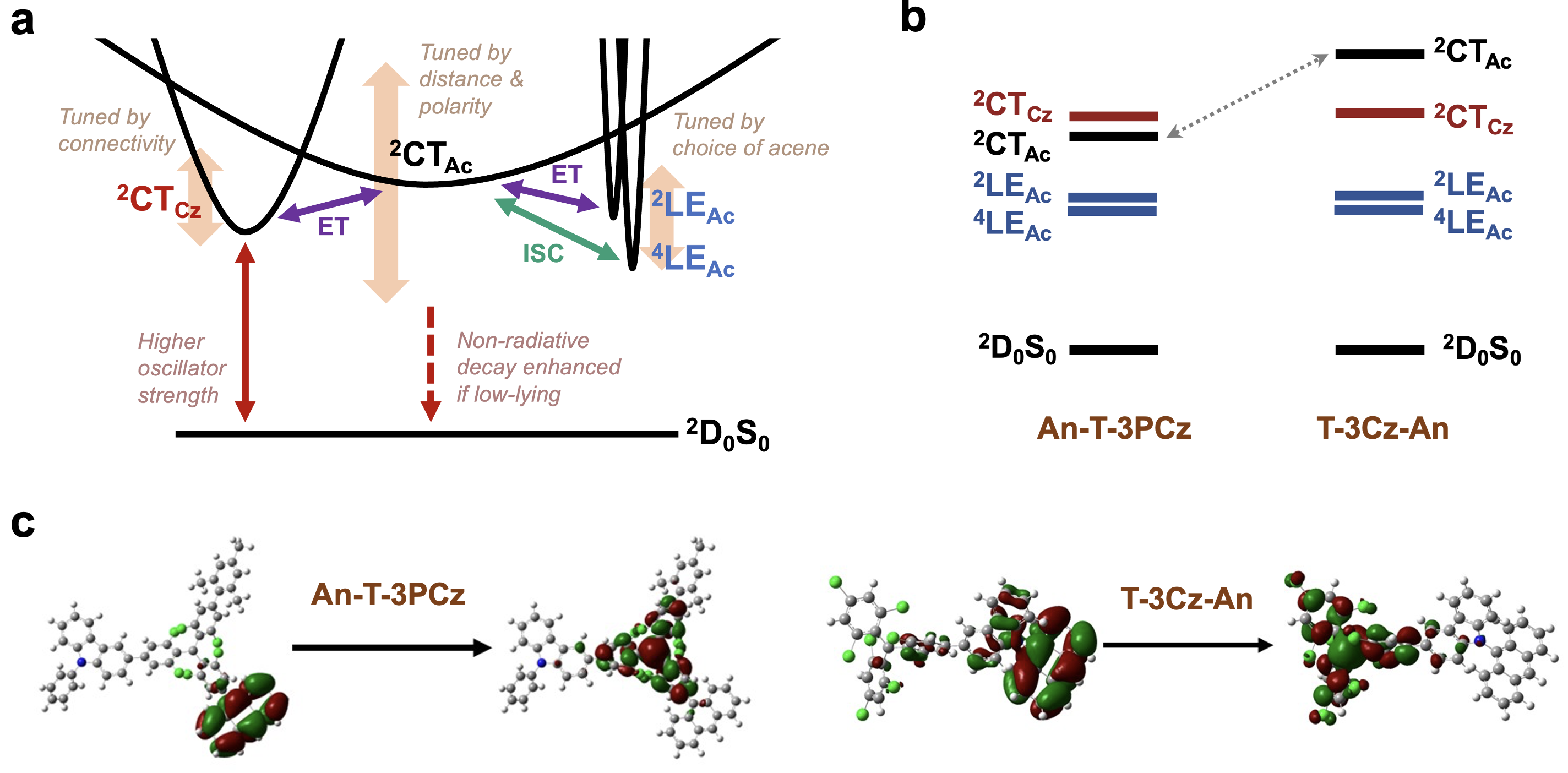}
  \caption{\textbf{Excited state landscape.} a) State diagram for a luminescent radical with an accessible quartet, with principal energy flow pathways. Spin conserving energy transfer (ET) as well as (R)ISC processes requiring a spin flip are indicated. The width of the parabolas indicates the energetic distribution of the corresponding states, and the length of the orange lines indicates the tuning range explored in the compounds presented here. b) Relative energetic ordering of key states in low dielectric constant environment. In An-T-3PCz and An-T-1Cz, the $^2$CT$_\text{Ac}$ state lies below the strongly absorbing and emissive $^2$CT$_\text{Cz}$, which limits emission yield. In T-3Cz-An and T-3Cz-Acr, the $^2$CT$_\text{Ac}$ does not act as a trap and assists in reversibly linking the emissive $^2$CT$_\text{Cz}$ and $^{4}$LE$_{\text{Ac}}$. c) Calculated NTOs for the $^2$CT$_\text{Ac}$ state. The extent of delocalisation increases from T-An, through An-T-3PCz, to T-3Cz-An. This delocalisation assists electronic coupling to other doublet states through wavefunction overlap.}
  \label{fgr:mech}
\end{figure}
 
The contrast between the reported compounds clearly shows that triplet-radical energy resonance alone does not guarantee efficient quartet formation in luminescent radicals. Covalent coupling of the chromophore and radical units inevitably introduces new states. In the case of radical-acene dyads, additional dark $^2$CT$_{\text{Ac}}$ and $^2$LE$_{\text{An}}$ states form. To minimise non-radiative decay via these states, tuning the energy difference between them and the bright $^2$CT$_{\text{Cz}}$ state is crucial (Fig. 4a). When the energy gap between $^2$CT$_{\text{Ac}}$ and $^2$LE$_{\text{An}}$ is small and the states are below the bright $^2$CT$_{\text{Cz}}$, non-radiative decay to the ground state can be substantial, especially if the states mix.\cite{FengLiBredas2023, Bredas2025-triplets} 

Conversely, efficient ISC via spin-orbit coupling requires states with different spatial wavefunctions. Therefore, the direct conversion of $^2$LE$_{\text{Ac}}$ to $^4$LE$_{\text{Ac}}$ is to first order forbidden. Since CT states offer a different spatial wave function, they can facilitate the spin-flip. For the $^2$CT$_{\text{Ac}}$ state to mediate (R)ISC, its energy needs to be above both the $^4$LE$_{\text{Ac}}$ and the emissive $^2$CT$_{\text{Cz}}$ state, but remain thermally accessible. If the $^2$CT$_{\text{Ac}}$ is significantly above the luminescent state, as in T-3Cz-Acr, the rate of ISC is reduced, leading to prompt emission kinetically competing with quartet generation. T-3Cz-An, with $^2$CT$_{\text{Ac}}$ only slightly above the luminescent state, outperforms T-3Cz-Acr despite acridine providing better energy alignment between $^2$CT$_{\text{Cz}}$ and $^4$LE$_{\text{Ac}}$ states. 

Thus, the TTM-acene electronic coupling strength plays a dual role in these systems. Firstly, it dictates the energetic stabilisation of $^2$CT$_{\text{Ac}}$ (Fig. 4b).  The small electron-hole separation in An-T-1Cz and An-T-3PCz lowers its energy below the $^2$CT$_{\text{Cz}}$. In contrast, the molecular structure of carbazole-bridged molecules circumvents the non-radiative decay through Coulombic interactions, by maximising electron-hole separation of $^2$CT$_{\text{Ac}}$. Secondly, it controls the energy transfer rates between local acene and charge transfer excitations.\cite{Lambert2012, Coropceanu2019} Indirect attachment of acene to TTM via carbazole slows the relaxation from $^2$LE$_{\text{Ac}}$ to the ground state, allowing ISC to kinetically outcompete non-radiative decay from the doublet states. Comparing An-T-3PCz and T-3Cz-An shows that molecular topology plays a crucial role in modulating the electronic rather than exchange coupling landscape. Due to the contribution of the TTM SOMO to the delocalised $\pi$-system, we expect strong exchange coupling can be maintained for distances exceeding 2 nm. 

More efficient RISC necessarily leads to a faster loss of spins on the $^4$LE$_{\text{Ac}}$ sublevel manifold. This represents a trade-off between luminescence yield and spin coherence times in our designs, and calls for compounds with efficient but relatively slow delayed emission. T-3Cz-An is the first reported design with a delayed emission lifetime over 1 µs, and its coherence times are similar to those of TEMPO-based quartet molecules.\cite{Maylander2024,Gorgon2024}

We summarize the additionally identified design criteria for this class of materials. Firstly, the radical fragment should not be an alternant hydrocarbon. Secondly, $^2$CT$_{\text{Ac}}$ should be high in energy, which can be achieved via spatial separation between the acene HOMO and radical SOMO, as well as a minimal energy gap between the SOMO and acene LUMO. Thirdly, strong exchange coupling can be obtained by minimizing the energy gap between the SOMO and the acene frontier orbitals as well as by linking the fragments at sites that have large amplitudes of SOMO, HOMO and LUMO. We exemplify these rules with T-3Cz-An, where a high luminescence yield was achieved by maximally decoupling the acene from the radical. 

This work presents a blueprint for compounds with microwave-addressable high spin states which spontaneously access a delayed, radiative decay channel. The demonstrated chemical control of electronic and exchange coupling marks a crucial step in the development of molecular quantum technologies.

\section{Materials and Methods}

\textit{Synthesis and characterization:} Detailed procedures are provided in the Supplementary Material.

\noindent
\textit{Steady-state spectroscopy:} Ultraviolet/visible absorption spectra were measured with a Shimadzu UV-2550 spectrometer. Fluorescence spectra and photoluminescence quantum yields were measured on a home-built setup consisting of cw laser diodes (Thorlabs) routed onto samples placed in an integrating sphere (Newport 819C-SL-5.3). Emission was injected into a spectrometer (Andor Kymera 328i) and onto a silicon CCD detector.

\noindent
\textit{Transient photoluminescence spectroscopy:} Time-resolved PL spectra were collected using an electrically-gated intensified CCD camera (AndoriStar DH740 CCI-010) after passing through a calibrated grating spectrometer (Andor SR303i). The samples were excited using pump pulses obtained from Orpheus-Lyra (Light Conversion).  Overlapping time regions were used to compose the decays at several constant gate widths. Temperature-dependent measurements were performed using a closed-circuit helium cryostat (Optistat Dry BL4, Oxford Instruments), compressor (HC-4E2, Sumitomo) and temperature controller (Mercury iTC, Oxford Instruments).

\noindent
\textit{Transient absorption spectroscopy:} ps TA experiments were conducted on either a home-built system pumped by a regenerative Ti:sapphire amplifier (Solstice Ace, Spectra-Physics) emitting pulses centered at 800 nm at a rate of 1 kHz, or a commercial HARPIA (Light Conversion) system driven by Pharos (Light Conversion) centered at 1035 nm at a rate of 10 kHz. For the home-built system, non-collinear optical parametric amplifiers (NOPA) were tuned to output a desired probe region, and a further NOPA tuned to provide narrowband pump pulses. Tunable pump pulses were generated in a Orpheus Neo (Light Conversion) unit. In both cases, one of the beams was optically delayed by a computer-controlled delay stage. The pump spectrum was filtered using appropriate band-pass filters to remove residual wavelengths, and its polarization set to magic angle relative to the probe. After passing through the sample, the probe was dispersed with a grating spectrometer and measured with either a Si or InGaAs detector array. ns TA experiments were performed on a home-built setup with excitation by the second harmonic (532 nm) of a Q-switched Nd:YVO$_4$ laser (Advanced Optical Technologies Ltd AOT-YVO-25QSPX, 500 Hz repetition rate) and the probe generated by LEUKOS Disco 1 supercontinuum laser (STM-1-UV, 1 kHz repetition rate). The time delay was varied using a Highland T560 delay generator. The probe was split with a 50:50 beam splitter to provide a pump-free reference beam. The transmitted probe and reference pulses were collected with a silicon dual-line array detector read out by a custom-built board (Stresing Entwicklungsbüro).

\noindent
\textit{Electron spin resonance:} X-band ESR was acquired with a Bruker Biospin E680 or E580 EleXSys spectrometer using a Bruker ER4118-MD5-W1 resonator (9.7 GHz) in an Oxford Instruments CF935 cryostat. For pulsed ESR, an Applied Systems Engineering (ASE) amplifier with saturated powers of 1.5 kW was used. Temperature was maintained with an ITC-503S temperature controller and a CF-935 helium flow cryostat (both Oxford Instruments). For laser-induced transient signals, photoexcitation was provided by Ekspla NT230 operating at a repetition rate of 50 Hz. Laser pulse energies used were 0.5 – 1 mJ, pulse lengths of 3 ns transmitted at ca. 40\% to the sample via the cryostat, microwave shield and resonator windows. A liquid-crystal depolarizer (DPP-25, ThorLabs) was placed in the laser path. Triggering of the laser and spectrometer involved synchronization with a Stanford Research Systems delay generator, DG645. Quadrature mixer detection was used in pulsed and continuous wave detection. Transient cw ESR spectra were simulated using EasySpin.

\noindent
\textit{Quantum Chemical Calculations:}
The doublet ground state structure of each molecule presented in this study was optimized at the Density Functional Theory (DFT) level in the unrestricted Kohn-Sham (UKS) formalism by using the $\omega$B97X-D exchange-correlation functional and the 6-31G(d,p) basis set for all the atomic species. The excited-state properties were computed by means of time-dependent (TD) DFT calculations within the Tamm-Dancoff approximation (TDA)\cite{HIRATA1999291}, where the unrestricted LC-$\omega$hPBE functional was used in conjunction with the Def2-TZVP basis set. For each molecule, the range-separation parameter $\omega$ was optimally tuned (OT) according to the ``gap-tuning'' procedure, as reported in earlier works\cite{Roi2012,Coropceanu2017effect}. To implicitly take into account the dielectric screening effects, the screened range-separated hybrid procedure (SRSH) was applied\cite{PhysRevB.88.081204}. In this approach, the sum of the parameters $\alpha + \beta$, defined within the DFT functional as the fraction of Hartree-Fock exchange included in the long-range domain, is set to be equal to $1/\varepsilon$, where $\alpha$ equals the amount of Hartree-Fock exchange applied in the short-range domain (usually set at 0.2), $\beta$ is an adjustable parameter varying as a function of $\varepsilon$, that is, the dielectric constant of the chosen solvent/environment. Calculations were carried out in toluene (TOL, $\varepsilon = 2.37$) and 2-methyltetrahydrofuran (MeTHF, $\varepsilon = 6.97$). We note that the unrestricted formalism does not allow for obtaining spin-contamination-free wave functions, and some excited states are inevitably affected by this issue. Hole-particle natural transition orbitals (NTOs) were produced to visually inspect the character of the relevant excitations for each group of radicals. Hole-electron separation distances and density overlaps were computed according to the detachment-attachment density formalism\cite{Monari2014toward}. All the calculations were performed using the \textsc{Gaussian16} package suite\cite{g16}.

\bibliography{achemso}

\begin{acknowledgement}

We thank Emrys Evans for discussions at the outset of this project. 

\noindent
\textbf{Funding:} European Research Council, European Union’s Horizon 2020 research and innovation program, grant no. 101020167 (LM, PM, RHF and SG); UKRI Engineering and Physical Sciences Research Council, grant no. EP/W017091/1 (HB) and grant nos. EP/V036408/1, EP/L011972/1 (WKM); National Natural Science Foundation of China, grant no. 51925303 (SY, WW, FL); John Fell Fund, grant nos. 0007019, 0010710 (WKM). DB and YO acknowledge funding by the Fonds de la Recherche Scientifique-FNRS under Grant n°T.0226.24 (OMLED). Computational resources were also provided by the ‘‘Consortium des Équipements de Calcul Intensif’’ (CÉCI), funded by the ‘‘Fonds de la Recherche Scientifiques de Belgique’’ (FRS-FNRS) under Grant No. 2.5020.11. GL thanks the Italian Ministry of University and Research for funding provided by the European Union-NextGenerationEU-PNRR, Missione 4, Componente 2, Linea di investimento 1.2. LM thanks Winton Programme and Harding Distinguished Postgraduate Scholarship for funding. PM and WZ have received funding from the European Union’s Horizon 2020 research and innovation program under the Marie Skłodowska-Curie grant agreements No. 891167 and No. 886066, respectively. PM thanks the Research Council of Finland (No. 363345). TJHH acknowledges a Royal Society University Research Fellowship URF\textbackslash R1\textbackslash 201502 and a startup grant from University College London. SG thanks Emmanuel College, Cambridge for a Research Fellowship.

\noindent
\textbf{Author contributions:} 
	Conceptualization: FL, HB, TJHH, RHF, SG;
	Methodology: LM, PM, SY, WW, JDG, GL, WKM, DB, YO, FL, HB, TJHH, SG;
	Investigation: LM, PM, SY, WW, JDG, GL, WZ, LB, SG;
	Visualization: LM, GL, SG;
	Supervision: DB, YO, FL, HB, TJHH, RHF, SG;
	Writing—original draft: LM, SG;
	Writing—review \& editing: LM, GL, DB, YO, TJHH, RHF, SG.

\noindent
\textbf{Competing interests:} Authors declare that they have no competing interests.\\
\noindent
\textbf{Data and materials availability:} Data are available from the University of Cambridge Apollo Repository at [URL to be added].

\end{acknowledgement}

\end{document}